\documentclass[twocolumn,showpacs,preprintnumbers,amsmath,amssymb]{revtex4}

\usepackage{graphicx}
\usepackage{dcolumn}
\usepackage{bm}

\begin{document}


\title{ $^{75}$As NMR study of the growth of paramagnetic-metal domains due to electron doping near the superconducting phase in LaFeAsO$_{1-x}$F$_{x}$
 }

 \author{ N. Fujiwara$^1$\footnote {Email:naoki@fujiwara.h.kyoto-u.ac.jp}, T. Nakano$^{1}$, Y. Kamihara$^{2}$, and H. Hosono$^{3}$
 }

\affiliation{$^1$Graduate School of Human and Environmental
Studies, Kyoto University, Yoshida-Nihonmatsu-cyo, Sakyo-ku, Kyoto
606-8501, Japan}

\affiliation {$^2$Department of Applied Physics $\&$ Physico-Informatics, Faculty of Science $\&$ Technology, Keio University, 3-14-1 Hiyoshi, Kohoku-ku, Yokohama, Kanagawa 223-8522, Japan }

\affiliation {$^3$Frontier Research Center (FRC), Tokyo
Institute of Technology, 4259 Nagatsuda, Midori-ku, Yokohama
226-8503, Japan}




\date{November 3 2010}


\begin{abstract}
We studied the electric and magnetic behavior near the phase boundary between antiferromagnetic (AF) and superconducting (SC) phases for a prototype of high-$T_c$ pnictides LaFeAsO$_{1-x}$F$_{x}$ by using nuclear magnetic resonance, and found that paramagnetic-metal (PM) domains segregate from AF domains. PM domains grow in size with increasing electron doping level and are accompanied by the onset of superconductivity, and thus application of pressure or increasing the doping level causes superconductivity. The existence of PM domains cannot be explained by the existing paradigm that focuses only on the relationship between superconductivity and antiferromagnetism. Based on orbital fluctuation theory, the existence of PM domains is evidence of the ferroquadrupole state.

\end{abstract}

\maketitle

\section{\label{sec:level1}Introduction}

In strongly correlated electron systems, including high-transition temperature ( high-$T_c$ ) superconductors, the electric and magnetic behavior at the phase boundary between antiferromagnetic ( AF ) and superconducting ( SC ) phases has attracted significant research interest. In iron-based high-$T_c$
pnictides, the AF state is a stripe-type spin-density-wave ( SDW ) state $^1$ arising from interband nesting between hole and electron pockets, and the relationship between AF and SC states is deeply connected with the pairing symmetry. Some theoretical investigations predict that SDW and SC order parameters are compatible near the phase boundary $^{2, 3}$, and the homogeneous coexistence of SDW and SC states is possible for superconductors with S$_{+-}$ symmetry $^{4, 5}$. In fact, this coexistence is
experimentally suggested for compounds that exhibit the crossover regime such as
Ba(Fe$_{1-x}$Co$_x$)$_2$As$_2$ ( Ba122 series )$^{6, 7}$, which is a representative high-$T_c$ pnictide, or CaFe$_{1-x}$Co$_{x}$AsF ( Ca1111 series )$^8$. In LaFeAsO$_{1-x}$F$_{x}$ ( La1111 series )$^9$, which is a prototype of high-$T_c$ pnictides, SDW and SC phases are segregated in the phase diagram, although the crossover regime
apparently exists in other $R$1111 series ($R$ = Ce, Nd, Sm, etc.) in which the highest $T_c$ is for the Sm1111 series $^{10, 11}$. The conditions in which
phase segregation or homogeneous coexistence appear remain to be elucidated for a variety of pnictides. Starting from the La1111 series, an empirical and systematic understanding is possible, which should aid in addressing this question.
Figure 1 shows the electronic phase diagrams of the La1111, Ca1111 and Ba122 series $^{9, 12-17}$. In these phase diagrams, $x$ represents the electron doping level and the horizontal axis corresponds to uniaxial compression.  For powder samples, uniaxial compression is attainable not by mechanical approaches, but by chemical approaches. The Ba122 and Ca1111 series are good examples of what can be attained by chemical approaches: the distance between FeAs planes
is 0.65 and 0.8593 nm for the Ba122 $^{18, 19}$ and Ca1111 $^{16}$ series, respectively, while that for the La1111 series is 0.8739 nm $^9$. With increasing uniaxial compression, (i.e. decreasing the distance between FeAs planes), the overlap between AF and SC phases increases, although $T_c$ remains unchanged. The homogeneous coexistence was observed for the compounds along the horizontal axis, implying the possibility of S$_{+-}$ symmetry $^{4, 5}$. The vertical axis in Fig. 1
corresponds to uniform compression. Unlike uniaxial compression, uniform compression is attainable by not only chemical approaches but also mechanical approaches. Other $R$1111 series ($R$ = Ce, Nd, Sm, etc.) have small lattice units and are equivalent to the La1111 series under hydrostatic
pressure $^{20}$. In fact, at 3.0 GPa, the La1111 series is nearly equivalent to
the Ce1111 series $^{21}$ in that the lattice constants are almost the same and the maximum
$T_c$ is about 40 K. Pressure application or rare earth replacement induces the overlap and shifts the maximum $T_c$ away from the phase boundary.
\begin{figure*}
\includegraphics{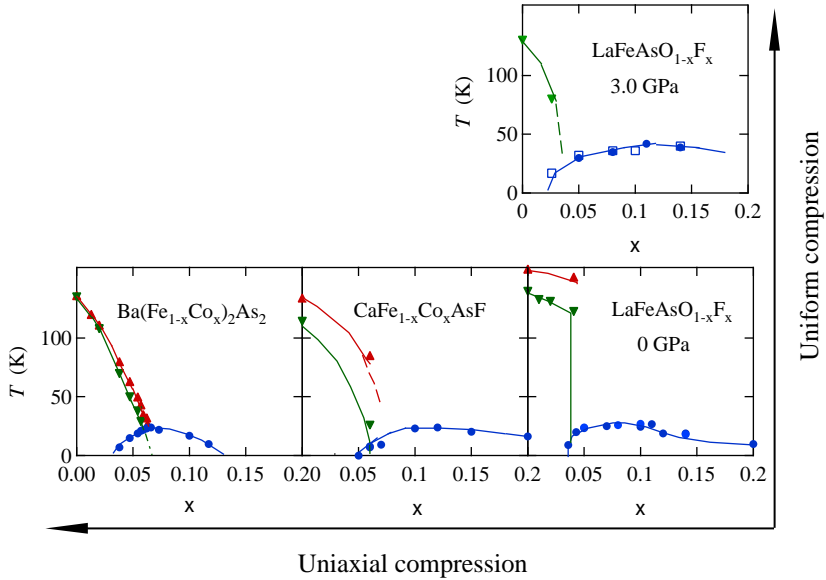}
\caption{\label{fig:epsart} (Color online) Phase diagrams of LaFeAsO$_{1-x}$F$_x$ ( La1111 ), CaFe$_{1-x}$Co$_{x}$AsF ( Ca1111 ) and Ba(Fe$_{1-x}$Co$_x$)$_2$As$_2$ ( Ba122 ). Uniform compression is equivalent to rare earth replacement in $R$FeAsO$_{1-x}$F$_x$ ( $R$1111 ), where $R$ = Ce, Nd, Sm, etc. Uniaxial compression along the crystal $c$ axis is attainable only by chemical approaches. The FeAs planes are more closely spaced for the Ba122 and Ca1111 series than for the La1111 series. Regular triangles and upside-down triangles represent structural ($T_S$), stripe-type antiferromagnetic ($T_N$) transition temperatures, respectively. Closed circles and open squares represent superconducting ($T_c$) transition temperatures determined from the resistivity and NMR measurements, respectively.}
\end{figure*}
\begin{figure}
\includegraphics{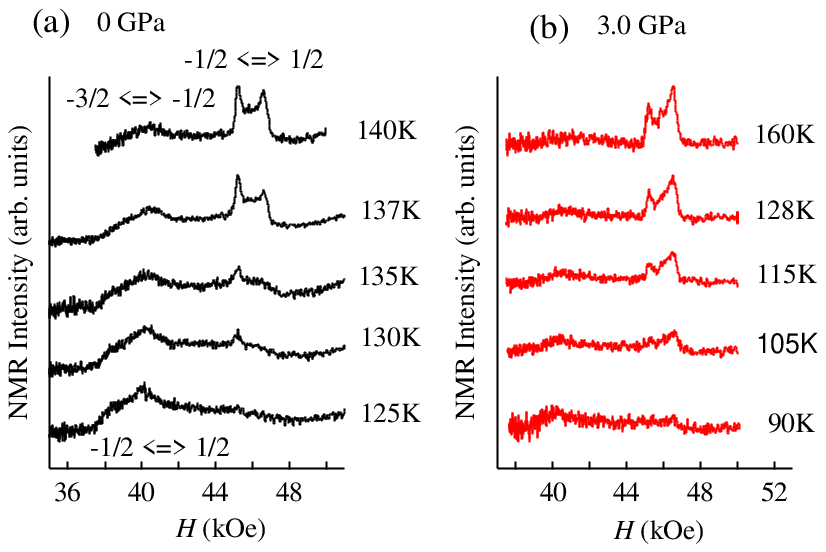}
\caption{\label{fig:epsart} (Color online)  $^{75}$As (I=$\frac{3}{2}$) NMR spectra measured at 35.1 MHz for undoped LaFeAsO. (a) NMR spectra at ambient pressure. Two sharp peaks correspond to the transition $I=-\frac{1}{2}\Leftrightarrow\frac{1}{2}$, and the broad low-field signal corresponds to the transition $I=-\frac{3}{2}\Leftrightarrow-\frac{1}{2}$. Double-peak structure appears due to the quadrupole interaction. At low temperatures, the signal corresponding to $I=-\frac{1}{2}\Leftrightarrow\frac{1}{2}$ is distributed to a wide field region because of the internal field arising from the ordered moments. (b) NMR spectra at 3.0 GPa. }
\end{figure}
For the La1111
series, the apparent overlap occurs upon application of either uniform or uniaxial pressure.  To determine whether homogeneous coexistence or phase segregation  occurs on a microscopic level for
uniform compression or rare earth replacement, we studied nuclear magnetic
resonance (NMR) spectra
under pressure $P$. We focus hereinafter on $^{75}$As ($I = \frac{3}{2}$) NMR spectra for the 2.6\% F-doped La1111 series because it is near the phase boundary.

\begin{figure}
\includegraphics{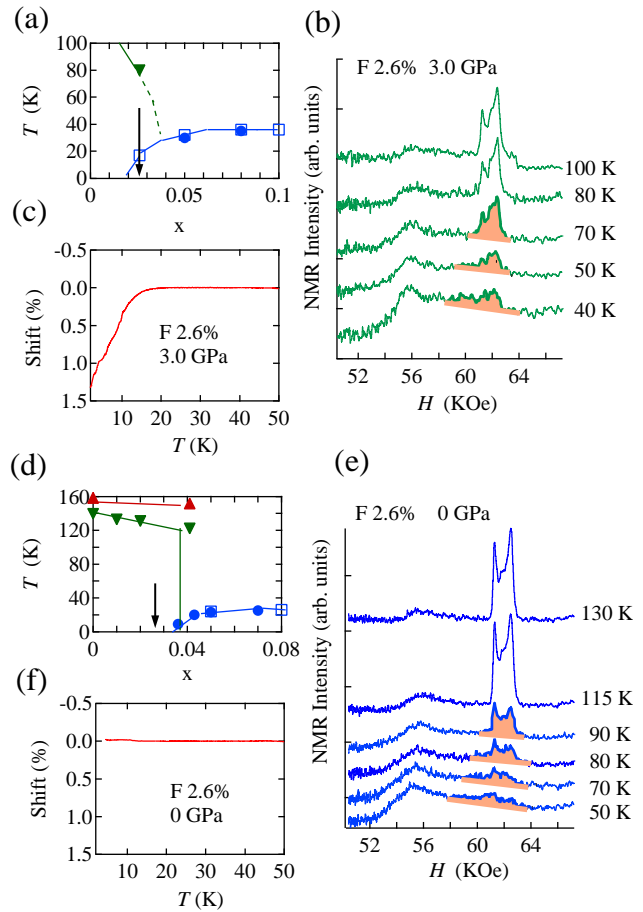}
\caption{\label{fig:epsart} (Color online)  (a) Phase diagram of LaFeAsO$_{1-x}$F$_{x}$ at 3.0 GPa expanded around the phase boundary. Closed circles and open squares represent $T_c$ determined from resistivity and NMR measurements, respectively. Some data were cited from Ref. 20 and 22. (b) $^{75}$As NMR spectra at 3.0 GPa measured at 45.1 MHz. Two peaks correspond to the transition $I =
-\frac{1}{2} \Leftrightarrow \frac{1}{2}$,  and the broad signal corresponds to the transition $I =
-\frac{3}{2}\Leftrightarrow -\frac{1}{2}$. The signals which originate from paramagnetic-metal phase below $T_N$ are highlighted in the spectra. (c) Detuning of the NMR tank circuit measured at the doping level of 2.6\% shown by an arrow in Fig. 3(a) under a pressure of 3.0 GPa. The bend indicates the onset of superconductivity. (d) Phase diagram of LaFeAsO$_{1-x}$F$_{x}$ at ambient pressure. (e) $^{75}$As NMR spectra at ambient pressure measured at 45.1 MHz. (e)Detuning of the NMR tank circuit measured at the doping level of 2.6\% shown by an arrow in Fig. 3(d) at ambient pressure. }
\end{figure}

\section{\label{sec:level1}Experimental results}

In the pulsed-NMR measurements, field-swept-NMR spectra were obtained from the spin-echo intensity after two coherent pulses. The relaxation rates ( 1/$T_1$ ) were measured by using the saturation-recovery method after a single pulse. A pressure of 3.0 GPa was applied by using a conventional clump-type pressure cell. Before discussing the 2.6\% doped samples, we show NMR spectra for the undoped samples measured at 35.1 MHz. Figures 2(a) and 2(b) show the spectra at ambient pressure and at 3.0 GPa, respectively. At high temperatures, the broad signal around 40 kOe and the sharper peak around 46 kOe correspond to the transitions $I = -\frac{3}{2} \Leftrightarrow -\frac{1}{2}$ and
$I = -\frac{1}{2} \Leftrightarrow \frac{1}{2}$, respectively. The sharper peak is split into a double-peak structure because of the second-order quadrupole effect. The signal corresponding to the transition, $I = \frac{3}{2} \Leftrightarrow \frac{1}{2}$ appeared at higher fields around 53 kOe, making a symmetric powder pattern. However, the latter broad signal overlaps the $^{139}$La NMR signal; therefore, throughout this article, we discuss only the broad lower-field signal and the double peaks. For the AF phase at low temperatures, the double-peak structure disappears and the signal corresponding to $I = -\frac{1}{2} \Leftrightarrow \frac{1}{2}$ is distributed in a wide field region because the internal fields arising from the AF ordering affect the resonance position. Intriguingly, the paramagnetic-metal ( PM ) state survives even below AF transition temperatures $T_N$ ($\sim$ 140 and 125 K at ambient pressure and 3.0 GPa, respectively), implying that a supercooling state is realized in these powder samples.

 A similar dependence on temperature $T$ is also seen for the 2.6\% F-doped La1111 series located near the phase boundary [ see Figs. 3(a) and 3(d) ]. Figures 3(b) and 3(e) show $^{75}$As NMR spectra at 3.0 GPa and at ambient pressure, respectively, measured at 45.1 MHz. At high temperatures, the NMR spectra are qualitatively the same as the undoped spectra. At low temperatures below $T_N$ ($\sim$ 100 and 80 K at ambient pressure and 3.0 GPa, respectively), the NMR spectra consist of two components: the broad signal originating from the AF state, and central bumps originating from the PM state as highlighted in Figs. 3(b) and 3(e). The central bumps are robust because they survive even at 10 K [ see Fig. 4(a) ], implying the occurrence of phase segregation or domain formation. At 3.0 GPa, PM domains become superconducting at low temperatures as seen from the detuning of the NMR tank circuit [see Fig. 3(c) ].  The detuning measured at 45.1 MHz indicates a $T_c$ value of 18K. Therefore, at 3.0 GPa, SC and AF phases become segregated on a microscopic level, although AF and SC phases apparently overlap in the phase diagram [ see Fig. 3(a) ]. The phase segregation between the AF and SC phases has been observed even for a doping level of 5.5\% by mean of muon-spin rotation ( $\mu$SR ), and pressure application causes an increase in the SC volume fraction against the AF volume fraction. $^{23}$  Interestingly, at ambient pressure the absence of the detuning as shown in Fig. 3 (f) indicates that phase segregation between AF and PM phases occurs, which is not expected from the existing phase diagram that indicates the AF phase at the 2.6\% doping level. The existence of the PM phase is the most important result from the present experiments and raises fundamental doubts about the existing paradigm that focuses only on the relationship between SC and AF states. Why does the PM phase occur at the doping level where the AF phase is expected in the existing diagram? The answer is deeply connected with the size of PM domains, which are not macroscopic but on the scale of several lattice units. At a doping level of 2.6\%, the PM domains are so small that some experimental techniques may not detect them. This assertion is supported by the fact that the relaxation rates ($1/T_1$) has a maximum at $T_N$ reflecting AF fluctuations from neighboring AF domains [ see Fig. 4(b) ]. Note that PM domains are due to neither a  nonuniform charge distribution nor some second phase. The former possibility is ruled out because domains with excess charge carriers would be not in a PM state but in a SC state. The latter possibility is also ruled out because $1/T_1$ would be free from AF fluctuations of neighboring domains.

\begin{figure}
\includegraphics{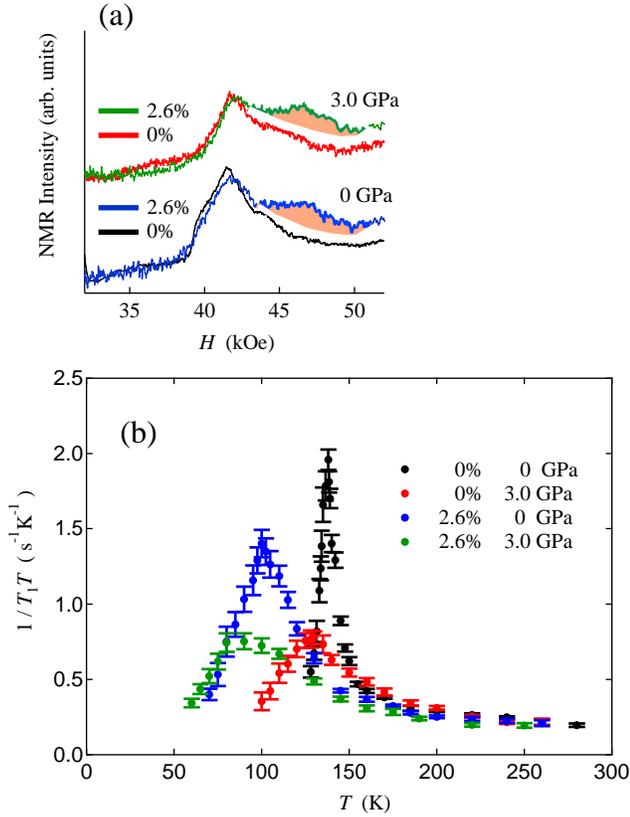}
\caption{\label{fig:wide} (Color online)  (a) $^{75}$As NMR spectra at 10 K measured at 35.1 MHz. Broad bumps at 3.0 GPa and at ambient pressure originate from superconducting and paramagnetic-metal phases, respectively. (b) Relaxation rates ($1/T_1$) measured at the lower-field peaks within the double-structure peaks corresponding to $I = -\frac{1}{2} \Leftrightarrow \frac{1}{2}$. $1/T_1$ has a maximum at $T_N$ reflecting antiferromagnetic fluctuations from neighboring AF domains. $1/T_1T$ for the undoped samples at ambient pressure were reported in Ref. 20, 24 and 25. }
\end{figure}

\begin{figure}
\includegraphics{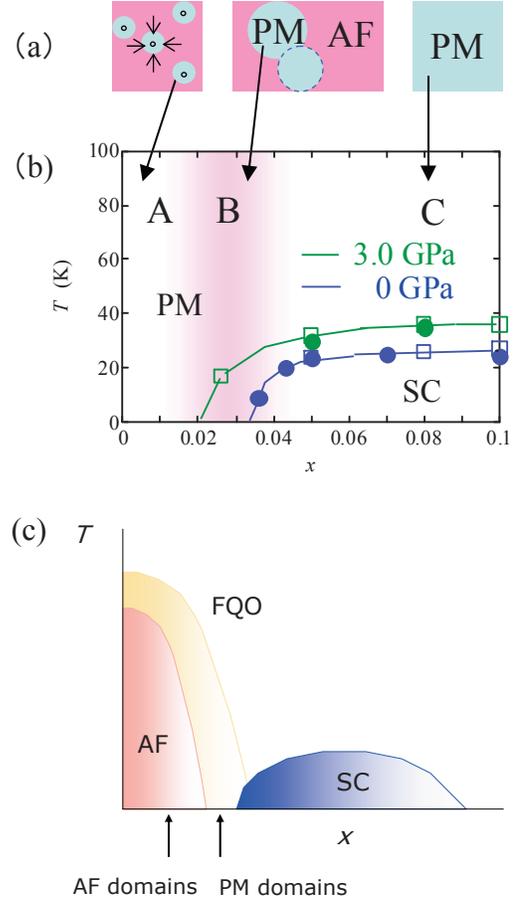}
\caption{\label{fig:wide} (Color online) (a) Schematic of paramagnetic-metal (PM) domains surrounded by antiferromagnetic domains. The three sequential drawings illustrate the growth of PM domains as the doping increases. PM domains in the underdoped regime [labeled A in panel(b)] shrinks in size with decreasing temperature, as indicated by the small arrows. (b) Doping-level ($x$) dependence of $T_c$ for PM domains. For 2.6\% F-doping level, PM domains is robust at ambient pressure, while they become superconducting at 3.0 GPa. The closed circles and open squares are the same with those in Figs. 3(a) and 3(d). (c) A phase diagram based on orbital fluctuation theory. The arrows indicate the doping levels of AF and PM domains in the central panel of Fig. 5(a).}
\end{figure}

\section{\label{sec:level1}Discussion}

The phase segregation between PM and AF phases cannot be explained by the existing paradigm: To understand the phenomenon together with the phase segregation between SC and AF phases at high pressure, two factors should be considered: (i) the growth of PM domains due to increasing electron doping level as illustrated in Fig. 5(a), and (ii) the location of the onset of superconductivity. PM domains undergo superconducting transition depending on $P$ and $x$ as shown in Fig. 5(b). In the underdoped regime [ regime A in Figs. 5(b) ], PM domains are maintained, even below $T_N$, as isolated seeds because of supercooling; however, they finally disappear at low temperatures. Therefore, the ground state is the AF state, which is consistent with existing observations. In the intermediate-doping regime [ regime B in Fig. 5(b) ], PM domains grow with increasing doping level and become robust even at low temperatures, causing phase segregation with AF domains. Whether the ground state is a SC or PM state depends on the location of the onset of superconductivity. The onset, which is at $\sim$ 3.5\% doping level at ambient pressure [ see Fig. 5(b) ], shifts to the underdoped regime upon applying pressure and crosses the 2.6\% doping level at 3.0 GPa. Therefore, two kinds of segregation are possible depending on pressure. In the intermediate-doping regime, applying pressure would not change the size of PM domains if one considers the NMR spectral intensity shown in Fig. 4(a). Taking account of a bulk volume fraction, PM domains would somehow link with neighboring PM domains, unlike in the underdoped regime. In the overdoped regime [ region C in Fig. 5(b) ], PM domains cover the entire system and exhibit superconducting properties at low temperatures, independent of pressure application. The PM state is free from AF, ( i.e., SDW ) ordering, implying that a factor other than interband nesting is crucial. Recently, a ferroquadrupole (FQ) ordering state between SDW and S$_{++}$ SC states has been suggested based on orbital fluctuation theory. $^{26, 27}$ A phase diagram based on the theory is shown in Fig. 5(c). AF and PM domains in the central panel of Fig. 5(a) correspond to different doping levels indicated by arrows in Fig. 5(c). PM domains at ambient pressure would be in the FQ ordering phase, while they become superconducting at 3.0 GPa because the SC phase boundary shifts to the underdoped regime by applying pressure. The electronic phase diagram shown in Fig. 5(c) allows us to reproduce the phase diagram in Fig. 1 because the volume fraction of AF domains is predominant around the phase boundary and therefore the contribution from AF domains is apparently emphasized for some experimental techniques.

\section{\label{sec:level1}Conclusion}

In conclusion, we have observed phase segregation between AF and PM domains at ambient pressure in the La1111 series by using $^{75}$As NMR. By increasing the electron doping level, we observed growth of PM domains accompanied by the onset of superconductivity. The PM state is independent of the AF ordering that arises from interband nesting, suggesting that the existing paradigm that focuses only on the relationship between superconductivity and antiferromagnetism is not valid. The FQ state predicted by orbital fluctuation theory is a leading candidate to explain the anomalous PM domains.

\textbf{Acknowledgements}

The NMR work is supported by a Grant-in-Aid (Grant No. KAKENHI 23340101) from the Ministry of Education, Science, and Culture, Japan. This work was supported in part by the JPSJ First Program. The authors would like to thank H. Kontani for discussion.







\end{document}